\documentclass[11pt]{article}
\usepackage[utf8]{inputenc}
\usepackage{amsmath,amssymb,bbm}
\usepackage{hyperref}
\hypersetup{
    colorlinks = true,
	citecolor=blue,
	linkcolor=blue
}
\usepackage{geometry}
\makeatletter
\renewcommand{\p@enumi}{Q}
\makeatother

\allowdisplaybreaks[4]

\usepackage{fourier}
\usepackage{stmaryrd}
\usepackage{mathtools}
\usepackage{newunicodechar}
\usepackage{subcaption}
\usepackage{algorithm}
\usepackage{algorithmicx}
\usepackage{algpseudocode}
\usepackage{bbm}
\usepackage{graphicx}[dvipdfmx]
\usepackage{mdframed}
\usepackage{natbib}
\usepackage{enumerate}
\usepackage{amsthm}
\theoremstyle{definition}

\newtheorem*{theorem*}{Theorem}
\newtheorem{proposition}{Proposition}

\newtheorem*{note*}{Note}

\newcommand{\argmax}{\mathop{\arg\max}}
\newcommand{\argmin}{\mathop{\arg\min}}
\newcommand{\bs}{\boldsymbol}
\newcommand{\diff}{\mathrm{d}}

\newcommand{\wht}{\widehat{\bs \theta}}
\newcommand{\wbt}{\overline{\bs \theta}}
\DeclarePairedDelimiter{\ip}{\langle}{\rangle}
\DeclarePairedDelimiter{\norm}{\|}{\|}
\DeclarePairedDelimiter{\stdnorm}{\llparenthesis}{\rrparenthesis}
\newcommand{\standardize}[1]{\mathsf{S}\left(#1\right)}
\newcommand{\covop}{\mathsf{X}}
\newcommand{\outop}{\mathsf{Y}}
\newcommand{\bridge}{\mathsf{L}}
\newcommand{\opt}{\min}
\newcommand{\pes}{\max}

\newcommand{\bft}{\boldsymbol{f_{\bs \theta}}}

\usepackage{authblk}

\title{An integrated perspective of robustness in regression \\ through the lens of the bias-variance trade-off}
\author[1,2]{Akifumi Okuno\thanks{okuno@ism.ac.jp}}
\affil[1]{Institute of Statistical Mathematics}
\affil[2]{RIKEN Center for Advanced Intelligence Project}
\date{\empty}

\begin{document}

\maketitle

\begin{abstract}
This paper presents an integrated perspective on robustness in regression. Specifically, we examine the relationship between traditional outlier-resistant robust estimation and robust optimization, which focuses on parameter estimation resistant to imaginary dataset-perturbations. While both are commonly regarded as robust methods, these concepts demonstrate a bias-variance trade-off, indicating that they follow roughly converse strategies.
\end{abstract}

\section{Introduction}

The concept of \emph{robustness} is of paramount importance across a variety of fields, particularly those involving practical statistical parameter estimation based on real-world observations. However, robust estimation techniques introduced in various methodologies aim to achieve different objectives, and each technique has been examined within individual frameworks. It is crucial to reexamine the purpose behind robust estimation and provide an integrated perspective across disciplinary boundaries. To facilitate this, this study initially classifies the goals of robust estimation methods into three categories: resistance to (1) outlier contamination (see, e.g., \citet{huber1981robust} and \citet{hampel1986robust}), (2) user-specified imaginary dataset-perturbation (see, e.g., \citet{ben2002robust} and \citet{biggio2013evasion}), and (3) model misspecification. Notably, (3) can be addressed using expressive models in certain cases; (3) will be discussed later but will not be the main focus. Therefore, this study primarily focuses on the following two categories within the context of linear regression:

\begin{enumerate}[{(1)}]
\item \emph{Outlier-resistance}. Outliers are data points that deviate significantly from the overall trend of the other observations in a dataset. Since the presence of outliers can affect statistical parameter estimation, potentially leading to unintended results, outlier-resistant estimation has been a focus for many decades~\citep{huber1981robust,hampel1986robust,maronna2006robust} mainly in the field of statistics. Originating from the works of \citet{tukey1960survey} and \citet{huber1964robust}, many outlier-resistant estimations are designed by modifying the loss function. Unlike the $L^2$ loss, which escalates the loss value for outliers deviating from the overall trend, traditional outlier-resistant estimations typically employ a loss function that less severely penalizes outliers. Interestingly, \citet{gannaz2007robust} and \citet{she2011outlier} reveal a relationship between outlier-resistant loss functions in line with \citet{huber1964robust} and the $L^2$ loss equipped with a slack variable, which is \emph{optimistically} estimated to negate the adverse effect of outliers.

\item \emph{Dataset-perturbation-resistance}. 
Robust optimization~\citep{ghaoui1997robust,ben-tal1998robust,ben2002robust,bertsimas2011theory}, whose prototype was proposed earlier by \citet{soyster1973technical}, considers optimizing the worst-case loss function among a user-specified imaginary dataset-perturbation. Robust optimization has been developed mainly in the field of numerical optimization. As the worst case is minimized, i.e., the loss function equipped with a slack variable is \emph{pessimistically} evaluated, the stability of the estimation is expected to improve, reducing the estimator variance. Robust optimization has also gained significant attention in recent years, particularly in the context of deep neural network training, where it is referred to as adversarial training~\citep{biggio2013evasion,szegedy2014intriguing,madry2018towards}.
\end{enumerate}

In summary, the statistical parameter estimation robust against (1) outlier contamination and (2) user-specified imaginary dataset-perturbation correspond to minimizing the loss function with optimistic and pessimistic perspectives, respectively. This study bridges these methods, and demonstrates their bias-variance trade-off. This trade-off indicates that they follow roughly converse strategies, where the optimistic one (i.e., outlier-robust) reduces the bias and the pessimistic one (i.e., robust optimization) reduces the variance typically in the presence of some impediments (e.g., outliers, dataset-perturbation), while both aim to reduce the overall generalization error. 

\begin{figure}[!ht]
\centering
\includegraphics[width=0.4\textwidth]{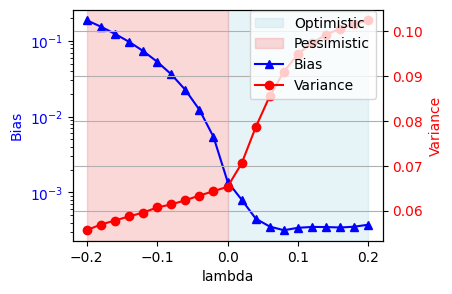}
\caption{Bias-variance trade-off between optimistic regression (i.e., outlier-robust estimation, where increasing $\lambda>0$ enhances the optimism) and pessimistic regression (i.e., robust optimization, where decreasing $\lambda<0$ enhances the pessimism). $\lambda=0$ represents OLS. This figure is a copy of Figure~\ref{fig:exp1_default}(\subref{subfig:default_(c)}); see Section~\ref{sec:experiments} for details.}
\end{figure}

In the following sections, we describe the related works and symbols. Section~\ref{sec:optimistic_pessimistic} then defines optimistic and pessimistic regression methods and provides their rationale. Section~\ref{sec:experiments} bridges these methods and demonstrates their bias-variance trade-off through numerical experiments. Finally, Section~\ref{sec:conclusion} concludes the paper. Source codes to reproduce the experimental results are available at \url{https://github.com/oknakfm/RBVT}.

\subsection{Related works}

While RO minimizes the worst-case loss function from a pessimistic perspective, minimizing the best-case scenario from an optimistic perspective is also known to yield better predictions. Optimism has been applied to classification~\citep{jinbo2004support}, multi-armed bandit problems~\citep{bubeck2012regret}, Bayesian optimization~\citep{srinivas2010gaussian,nguyen2019optimistic}, and clustering~\citep{okuno2022greedy,hattori2023finding}. On the other hand, pessimism has been introduced into asset allocation~\citep{tutuncu2004robust}, classification~\citep{xu2009robustness,takeda2013unified}, and clustering~\citep{vo2016robust} within the context of robust optimization.

Although technically slightly different, distributionally robust optimization~\citep{wiesemann2014distributionally,chen2020distributionally} considers the perturbation of the empirical distribution of the observations and can be regarded as a form of variance regularization~\citep{namkoong2017variance,duchi2019variance}. Within this framework, the bias-variance trade-off along with the perturbation degree has been theoretically investigated. See, e.g., \citet{gotoh2018robust,gotoh2021calibration}. While the bias-variance trade-off has been examined within each robustness framework, our study provides an integrated perspective across disciplinary boundaries, highlighting the roughly converse relationship between outlier-resistant robust estimation and robust optimization.


\subsection{Symbols and Notation}

For a user-specified exponent $q>0$, the $q$-norm and the standardized $q$-norm of a vector $\bs z=(z_1,z_2,\ldots,z_m) \in \mathbb{R}^m$, where $m \in \mathbb{N}$, are defined respectively as:
\[
\norm{\bs z}_q
=
\left\{\sum_{i=1}^{m} |z_i|^q\right\}^{1/q},
\quad
\stdnorm{\bs z}_q = \left\{\frac{1}{m} \sum_{i=1}^{m} |z_i|^q\right\}^{1/q}.
\]
By taking the limit $q \to \infty$, we obtain $\norm{\bs z}_{\infty}=\stdnorm{\bs z}_{\infty}=\max_j |z_j|$. The symbols $\bs \Delta,\bs \gamma$ represent a $n \times d$ matrix and a $n$-dimensional vector that denote the perturbations in the observed design matrix and outcomes, respectively. More specifically, $\bs \Delta=(\delta_{ij}) \in \mathbb{R}^{n \times d}$ where $\bs \delta_i=(\delta_{i1},\delta_{i2},\ldots,\delta_{id}) \in \mathbb{R}^d,\tilde{\bs \delta}_j=(\delta_{1j},\delta_{2j},\ldots,\delta_{nj}) \in \mathbb{R}^n$ denote its $i$th row and $j$th column, and $\bs \gamma = (\gamma_1,\gamma_2,\ldots,\gamma_n) \in \mathbb{R}^n$ with $\gamma_i$ representing the $i$th entry. Using the norm $\norm{\cdot}_q$ defined for any vector, a standardized ($\alpha,\beta$)-norm $\stdnorm{\cdot}_{\alpha,\beta}$ with hyperparameters $\alpha,\beta>0$ is defined for any matrix $\bs \Delta$ as:
\begin{align}
    \stdnorm{\bs \Delta}_{\alpha,\beta}
    :=
    \stdnorm{\, \left(
        \stdnorm{\bs \delta_1}_{\beta},
        \stdnorm{\bs \delta_2}_{\beta},
        \ldots,
        \stdnorm{\bs \delta_n}_{\beta}
        \right) \,
    }_{\alpha}.
    \label{eq:rs-norm}
\end{align}
We obtain the identity $\stdnorm{\bs \Delta}_{\alpha,\alpha}=(\sum_{i,j} \delta_{ij}^{\alpha}/nd)^{1/\alpha}$ for any $\alpha>0$, where $\stdnorm{\bs \Delta}_{2,2}=(\sum_{i,j} \delta_{ij}^2 /nd)^{1/2}$ is particularly referred to as a standardized Frobenius norm. The indicator function $\mathbbm{1}_{\mathcal{A}}$ assumes a value of $1$ if the event $\mathcal{A}$ occurs and $0$ otherwise. The inner product $\ip{\bs z,\bs z'}=\sum_{i=1}^{m} z_i z_i'$ is defined for any vectors $\bs z,\bs z' \in \mathbb{R}^m$, and similarly, 
\[
\ip{\bs \Delta,\bs \Delta'}=\sum_{i=1}^{n}\sum_{j=1}^{d}\delta_{ij}\delta_{ij}'
\]
is defined for matrices $\bs \Delta,\bs \Delta' \in \mathbb{R}^{n \times d}$. The zero vector of dimension $d$ is denoted by $\bs 0_d$, and $\bs I_d$ denotes the $d \times d$ identity matrix. The normal distribution with mean $\bs \mu$ and variance-covariance matrix $\bs \Sigma$ is denoted by $N_d(\bs \mu,\bs \Sigma)$. For $d=1$, it is simply written as $N(\mu,\sigma^2)$. The relation $L_1(\bs \theta) \propto L_2(\bs \theta)$ indicates that the functions $L_1$ and $L_2$ are proportional, i.e., there exists $\alpha>0$ such that $L_1(\bs \theta)=\alpha L_2(\bs \theta)$ for any $\bs \theta$.

\section{Optimistic Regression and Pessimistic Regression}
\label{sec:optimistic_pessimistic}

In this section, we first outline the framework of ordinary least squares (OLS) regression in Section~\ref{sec:OLS}. We then introduce the concepts of optimistic and pessimistic regression in Sections~\ref{sec:optimistic} and \ref{sec:pessimistic}, respectively. As discussed in each section, optimistic regression is closely associated with outlier-resistant robust estimation. Conversely, pessimistic regression, which is considered robust optimization applied to regression problems, can be interpreted as a regularization of the regression model. Theories of optimistic and pessimistic methods are presented in this section, and they are formally integrated in the next Section~\ref{sec:experiments} to demonstrate their bias-variance trade-off.

\subsection{OLS regression}
\label{sec:OLS}

Let $n,d \in \mathbb{N}$ represent the sample size and the dimension of covariates, respectively. $\{(\bs x_i,y_i)\}_{i=1}^{n}$ consists of $n$ pairs of observed covariates and outcomes, where $\bs X=(\bs x_1^{\top},\bs x_2^{\top},\ldots,\bs x_n^{\top})^{\top} \in \mathbb{R}^{n \times d}$ is the observed design matrix and $\bs y=(y_1,y_2,\ldots,y_n)$ is the vector of observed outcomes. The function $f_{\bs \theta}:\mathbb{R}^d \to \mathbb{R}$ is parameterized by $\bs \theta \in \mathbb{R}^d$ and is to be estimated. The vector $\bft(\bs X)=(f_{\bs \theta}(\bs x_1),f_{\bs \theta}(\bs x_2),\ldots,f_{\bs \theta}(\bs x_n))$ concatenates the application of $f_{\bs \theta}$ to each row vector $\bs x_i$ of $\bs X$. 

The minimization problem
\begin{align}
    \wht = \argmin_{\bs \theta \in \mathbb{R}^d}
    \stdnorm{\bs y-\bft(\bs X)}_2^2
\label{eq:OLS}
\end{align}
defines OLS regression, which also aligns with the maximum likelihood estimator assuming the probabilistic model $\bs y \mid \bs X \sim N_n(\bft(\bs X),\sigma^2 \bs I_n)$. If the linear model $f_{\bs \theta}(\bs x)=\ip{\bs x,\bs \theta}$ is employed and the design matrix $\bs X$ is of full rank, the estimator \eqref{eq:OLS} simplifies to $\wht=(\bs X^{\top}\bs X)^{-1}\bs X^{\top}\bs y$.

\subsection{Optimistic regression}
\label{sec:optimistic}

This section introduces the concept of optimistic regression. As will be described in this section, optimistic regression can be readily interpreted through maximum likelihood estimation employing specific probabilistic models. Additionally, it also exhibits a close relationship with outlier-resistant robust regression, emphasizing its utility in handling data anomalies effectively.

Here, we define two types of optimistic loss functions: \emph{covariate-optimistic} and \emph{outcome-optimistic} loss functions: 
\begin{align*}
    \covop_{\alpha,\beta,\gamma,\tau}^{\opt}
    (\bs y,\bft(\bs X))
    &:=
    \min_{\bs \Delta \in \mathbb{R}^{n \times d}}
    \left\{ 
    \stdnorm{\bs y-\bft(\bs X+\bs \Delta)}_2^2
    +
    \tau^{-1} \stdnorm{\bs \Delta}_{\alpha,\beta}^{\gamma}
    \right\}, \\
    \outop_{\alpha,\gamma,\tau}^{\opt}(\bs y,\bft(\bs X))
    &:=
    \min_{\bs \mu \in \mathbb{R}^n}
    \left\{ 
    \stdnorm{\bs y+\bs \mu-\bft(\bs X)}_2^2
    +
    \tau^{-1} \stdnorm{\bs \mu}_{\alpha}^{\gamma}
    \right\}, 
\end{align*}
where $\alpha,\beta,\gamma> 0, \tau \ge 0$ are user-specified hyperparameters. Consider a case $\tau \searrow 0$: the entries of $\bs \Delta$ and $\bs \mu$ approach zero, causing these loss functions to reduce to the OLS loss function $\stdnorm{\bs y-\bft(\bs X)}_2^2$. From a statistical perspective, these two optimistic methods can be understood as follows.

\subsection*{Covariate-optimistic regression} 

The covariate-optimistic loss function can be interpreted as a (partially) maximized log-likelihood for an errors-in-variables model. To elucidate this, consider a simple errors-in-variable model:
\begin{align*}
    \bs y \mid \widetilde{\bs X},\bs \theta \sim N_n\left(\bft(\widetilde{\bs X}), \frac{n}{2}\bs I_n\right),
    \quad 
    \bs X \mid \widetilde{\bs X} \sim E_{\alpha,\beta,\gamma,\tau}(\widetilde{\bs X}),
\end{align*}
where $E_{\alpha,\beta,\gamma,\tau}$ represents a covariate distribution whose density function is proportional to $\exp(-\frac{1}{\tau}\stdnorm{\bs X - \widetilde{\bs X}}_{\alpha,\beta}^{\gamma})$. By considering a change-of-variable $\tilde{\bs X}=\bs X+\bs \Delta$, it becomes clear that the negative sign of the partially maximized log-likelihood coincides with the covariate-optimistic loss function:
\begin{align}
-\max_{\widetilde{\bs X} \in \mathbb{R}^{n \times d}} \log \{\mathbb{P}(\bs y\mid \widetilde{\bs X},\bs \theta)\mathbb{P}(\bs X \mid \widetilde{\bs X})\}
=
\covop^{\opt}_{\alpha,\beta,\gamma,\tau}(\bs y,\bft(\bs X)).
\end{align}

Thus, the covariate-optimistic regression that minimizes the above optimistic loss function is consistent with the maximum likelihood estimation for the errors-in-variables model:
\[
    \argmin_{\bs \theta \in \mathbb{R}^d} \covop^{\opt}_{\alpha,\beta,\gamma,\tau}(\bs y,\bft(\bs X))
    =
    \argmax_{\bs \theta \in \mathbb{R}^d} \max_{\widetilde{\bs X} \in \mathbb{R}^{n \times d}} \log \{\mathbb{P}(\bs y\mid \widetilde{\bs X},\bs \theta)\mathbb{P}(\bs X \mid \widetilde{\bs X})\}.
\]
In the limit as $\tau \searrow 0$, i.e., $\bs X$ coincides with $\tilde{\bs X}$ with probability $1$, covariate-optimistic regression simplifies to OLS regression. 
Covariate-optimistic regression is also referred to as min-min regression in \citet{bertsimas2017trimmed}. 

Particularly noteworthy, specifying $\alpha=\beta=\gamma=2$ transforms $E_{\alpha,\beta,\gamma,\tau}$ into a Gaussian covariate distribution. Known as Deming regression under these parameters, this method aligns with statistical consistency for the appropriately chosen hyperparameter $\tau>0$, as described by \cite{kummell1879reduction} and \citet{deming1943statistical}. This regression is a specific instance of orthogonal regression or total least squares~(see, e.g., \citet{adcock1878problem}). For comprehensive surveys of other errors-in-variables regression approaches, including instrumental methods, see \citet{gillard2010overview}. 

We also note that \citet{bertsimas2017trimmed} regards the above covariate-optimistic regression with $\alpha=\beta=2$ and the linear model $f_{\bs \theta}(\bs x)=\langle \bs x,\bs \theta\rangle$ is like the least trimmed squares regression~\citep{rousseeuw1987robust}. We think that the least trimmed squares, that minimizes $\sum_{i=j+1}^{n}r_i(\bs \theta)^2$ where $r_i(\bs \theta)=|y_i-\langle \bs x_i,\bs \theta\rangle|$ are absolute residuals and $r_{(i)}(\bs \theta)$ are the sorted residuals, is more similar to the outcome-optimistic regression. \citet{bertsimas2017trimmed} also incorporates the trimmed-Lasso penalty on the parameter $\bs \theta$ into the covariate-optimistic regression.

\subsection*{Outcome-optimistic regression} 

Similarly to covariate-optimistic regression, outcome-optimistic regression can also be framed with a probabilistic model:
\[
    \bs y \mid \bs X,\bs \mu,\bs \theta
    \sim 
    N_n\left(\bft(\bs X)+\bs \mu,\frac{n}{2}\bs I_n\right), 
    \quad 
    \bs \mu \sim F_{\alpha,\gamma,\tau},
\]
where $F_{\alpha,\gamma,\tau}$ denotes a distribution whose density function is proportional to $\exp\left(-\frac{1}{\tau}\stdnorm{\bs \mu}_{\alpha}^{\gamma}\right)$. The negative sign of the partially maximized log-likelihood aligns with the outcome-optimistic loss function, as follows:
\begin{align}
-\max_{\bs \mu \in \mathbb{R}^{n}} \, 
\log\{\mathbb{P}(\bs y \mid \bs X,\bs \mu,\bs \theta)\mathbb{P}(\bs \mu)\}
=
\outop^{\opt}_{\alpha,\gamma,\tau}(\bs y,\bft(\bs X)).
\end{align}
Therefore, minimizing the outcome-optimistic loss function corresponds to considering an additional (contaminated) vector $\bs \mu$ in the observed outcome $\bs y$. This type of contaminated vector is typically considered in the field of outlier-resistant robust statistics. 

In line with \citet{gannaz2007robust}, several studies have demonstrated a direct equivalence between the outcome-optimistic loss function and outlier-resistant robust loss functions. Refer to the following Proposition~\ref{prop:gannaz}: although \citet{gannaz2007robust} initially considered only the linear regression function $f_{\bs \theta}(\bs x)=\ip{\bs x,\bs \theta}$, the proof can be straightforwardly generalized to encompass arbitrary nonlinear function $f_{\bs \theta}$. This adaptation underscores the versatility of the outcome-optimistic regression approach in addressing a broad spectrum of statistical modeling scenarios. 

\begin{proposition}[A nonlinear extension of \citet{gannaz2007robust}]
\label{prop:gannaz}
It holds for the Huber's function $\rho_{\eta}(u)=(u^2/2) \mathbbm{1}_{|u| \le \eta}+\{\eta|u|-\eta^2/2\}\mathbbm{1}_{|u|>\eta}$ that $\outop^{\opt}_{1,1,\tau}(\bs y,\bft(\bs X)) \propto \sum_{i=1}^{n}\rho_{n/2d\tau}(y_i-f_{\bs \theta}(\bs x_i))+C$, 
where $C \in \mathbb{R}$ is a constant independent of $\bs \theta$. 
\end{proposition}

As has been extensively discussed in the field of robust statistics, statistical estimation using Huber's loss function $\rho_{\eta}(\cdot)$ is known for its resistance to outliers. Seminal works by \citet{huber1964robust}, \citet{huber1981robust}, \citet{hampel1986robust} and \citet{maronna2006robust} describe the details. Therefore, outcome-optimistic regression can be considered as a form of outlier-resistant robust regression. 

Expanding on this line of research, \citet{she2011outlier} generalizes the framework outlined in Proposition~\ref{prop:gannaz} by replacing the penalization term $\tau^{-1}\stdnorm{\bs \mu}_1$ used in \citet{gannaz2007robust} with $\sum_{i=1}^{n}P(\mu_i)$ for a more general function $P$. \citet{katayama2017sparse} extends the work of \citet{she2011outlier} into high-dimensional settings by incorporating a sparse penalty of the parameter vector $\bs \theta$, thereby enhancing the model's applicability and effectiveness in scenarios with high-dimensional data.

\subsection{Pessimistic Regression}
\label{sec:pessimistic}

This section introduces the concept of pessimistic regression. In contrast to optimistic regression, which focuses on the best-case loss function concerning covariate and outcome perturbations, pessimistic regression contemplates the worst-case loss function. Optimizing this pessimistic loss function aligns with the principles of robust optimization and is recognized as adversarial training in contemporary machine learning literature. This approach underscores the strategic defense against potential worst-case scenarios in predictive modeling.

Herein, we define the pessimistic loss function. 
It is important to note that merely replacing the $\min$ function and the penalty term in the optimistic loss function $\covop^{\opt}_{\alpha,\beta,\tau}$ with a $\max$ function and the reversed-sign penalty can result in the infimum. To address this issue, we define the \emph{covariate-pessimistic} and \emph{outcome-pessimistic} loss functions via a constrained optimization problem:
\begin{align*}
    \covop_{\alpha,\beta,\gamma,\tau}^{\pes}(\bs y,\bft(\bs X))
    &:=
    \max_{\stdnorm{\bs \Delta}_{\alpha,\beta}^{\gamma} \le \tau}
    \stdnorm{\bs y-\bft(\bs X+\bs \Delta)}_2^2, \\
    \outop_{\alpha,\gamma,\tau}^{\pes}(\bs y,\bft(\bs X))
    &:=
    \max_{\stdnorm{\bs \mu}_{\alpha}^{\gamma} \le \tau}
    \stdnorm{\bs y+\bs \mu-\bft(\bs X)}_2^2.
\end{align*}
Notably, both these loss functions converge to the OLS loss function as $\tau \searrow 0$.

\subsubsection*{Covariate-pessimistic regression}
In contrast to optimistic regression, which can be viewed as a maximum likelihood estimation under dataset-perturbation, the covariate-pessimistic loss function integrates parameter penalization into the loss function implicitly. 

To elucidate the effect of this penalization, consider a single step of the gradient descent initiated with $\bs \Delta^{(0)}=\bs O$. This single-step update is described as
\[
    \bs \Delta^{(1)}
    =
    \bs \Delta^{(0)}
    +
    \tau^{1/\gamma}
    \standardize{\bs \Omega},
\]
where the gradient and the operator for standardization are respectively defined as 
\[
    \bs \Omega:=\frac{\partial \stdnorm{\bs y-\bft(\bs X+\bs \Delta)}_2^2}{\partial \bs \Delta} \Bigg|_{\bs \Delta=\bs O},
    \quad 
    \standardize{\bs \Omega}:=
    \begin{cases}
    \bs \Omega/\stdnorm{\bs \Omega}_{\alpha,\beta} & (\bs \Omega \ne \bs O) \\ 
    \bs O & (\bs \Omega = \bs O) \\
    \end{cases}.
\]
This update is expected to roughly approximate the maximizer of the loss function $\stdnorm{\bs y-\bft(\bs X+\bs \Delta)}_2^2$  under the assumption that the $\tau>0$ is sufficiently small; it satisfies the constraint $\stdnorm{\bs \Delta^{(1)}}_{\alpha,\beta}^{\gamma} \le \tau$. Consequently, we derive:
\begin{align*}
    \covop^{\pes}_{\alpha,\beta,\gamma,\tau}(\bs y,\bft(\bs X))
    \, \approx \,
    \stdnorm{\bs y-\bft(\bs X+\bs \Delta^{(1)})}_2^2 
    \, \overset{(\star)}{\approx} \,
    \stdnorm{\bs y-\bft(\bs X)}_2^2
    +
    \ip{\bs \Omega,\bs \Delta^{(1)}}
\end{align*}
where $\approx$ denotes approximation and $(\star)$ is a linear approximation. Specifically, in the simple case where $\alpha=\beta=2$, the last term in the formula reduces to
\begin{align}
    \ip{\bs \Omega,\bs \Delta^{(1)}}
    &=
    \tau^{1/\gamma}
    \ip{\bs \Omega,\standardize{\bs \Omega}}
    =
    \tau^{1/\gamma}
    \frac{\ip{\bs \Omega,\bs \Omega}}{\stdnorm{\bs \Omega}_{2,2}}
    =
    \tau^{1/\gamma}
    \frac{nd\stdnorm{\bs \Omega}_{2,2}^{2}}{\stdnorm{\bs \Omega}_{2,2}}
    =
    \tau^{1/\gamma}
    nd
    \stdnorm{\bs \Omega}_{2,2} \nonumber \\
    &=
    2
    \tau^{1/\gamma}
    d
    \left\{
    \frac{1}{n}
    \sum_{i=1}^{n}
    e_{\bs \theta,i}^2 
        \stdnorm{\frac{\partial f_{\bs \theta}(\bs x_i)}{\partial \bs x_i}}_2^2
    \right\}^{1/2}
    \label{eq:pen_term}
\end{align}
for $\bs \Omega \ne \bs O$ (and $0$ if $\bs \Omega=\bs O$), with the residual $e_{\bs \theta,i}:=y_i-f_{\bs \theta}(\bs x_i)$. Since $e_{\bs \theta,i}^2$ is always non-negative, the term \eqref{eq:pen_term} is interpreted to represent penalization for the variation in the function $f_{\bs \theta}$ (which is represented as derivatives concerning the function input). See, e.g., \citet{okuno2023stochastic} for the variation regularization for non-linear models. Even when the assumption $\alpha=\beta=2$ is not met, the term $\ip{\bs \Omega,\bs \Delta^{(1)}}$is calculated as $\tau^{1/\gamma}nd\stdnorm{\bs \Omega}_{2,2}^{2}/\stdnorm{\bs \Omega}_{\alpha,\beta}$. Given that the numerator represents the second-order variation and the denominator the first-order, this term is still expected to work as a penalization for variation in the prediction function $f_{\bs \theta}$. 

Notably, recent work by \citet{imaizumi2023sup} has considered adversarial training (i.e., covariate-pessimistic optimization) in the context of deep neural networks with decreasing $\tau=\tau_n$ ($n \to \infty$), demonstrating that the trained network achieves minimax optimality in terms of uniform convergence. Although the concepts and technical details differ, our explanation shown above further reinforces the assertions made by \citet{imaizumi2023sup}. 

\bigskip
Here, let us revisit the simple case $\alpha=\beta=2$. Considering the linear case $f_{\bs \theta}(\bs x)=\ip{\bs x,\bs \theta}$ and assuming $\bs \Omega \ne \bs O$, the interpretation of term \eqref{eq:pen_term} simplifies. It effectively acts as parameter regularization:
\[
\eqref{eq:pen_term}
=
\kappa_{\tau} \cdot \stdnorm{\bs \theta}_2
\]
with the coefficient $\kappa_{\tau}=2\tau^{1/\gamma}d\stdnorm{\bs y-\bs X\bs \theta}_2$. While this discussion assumes that $\tau>0$ is sufficiently small (whereby the coefficient $\kappa_{\tau}$ is also small), \citet{ribeiro2023regularization} provides similar insights to any $\tau>0$ not limited to $\tau \approx 0$, proving the equivalence between covariate-pessimistic regression and parameter regularization under broader conditions. Refer to the following Proposition for details: 

\begin{proposition}[\citet{ribeiro2023regularization} Proposition~1]
\label{prop:ribeiro}
    It holds for any $\beta,\beta_*>0$ satisfying $\beta^{-1}+\beta_*^{-1}=1$ that 
    $\covop^{\pes}_{\infty,\beta,1,\tau/d}(\bs y,\bft(\bs X))=n^{-1}\sum_{i=1}^{n}\left\{
        |y_i-\ip{\bs x_i,\bs \theta}| 
        +
        \tau \stdnorm{\bs \theta}_{\beta_*}
    \right\}^2$.
\end{proposition}

Proof is simply obtained by substituting $\delta=\tau d^{1/\beta}$ and $\norm{\cdot}=\norm{\cdot}_{\beta}$ into the proof of \citet{ribeiro2023regularization} Proposition~1. Notably, specifying $\beta \to \infty$ leads to $\beta_* \to 1$, and it indicates that the covariate-pessimistic loss function, computed with the sup-norm constraint $\stdnorm{\bs \Delta}_{\infty,\infty}=\max_{i,j}|\delta_{i,j}| \le \tau$ (i.e., $\alpha=\beta=\infty$), simplifies to the Lasso-like loss function: $n^{-1}\sum_{i=1}^{n}\{|y_i-\ip{\bs x_i,\bs \theta}|+\tau\stdnorm{\bs \theta}_1\}^2$. \citet{ribeiro2023overparameterized} further evaluates the generalization error for overparameterized linear regression using adversarial training, i.e., covariate-pessimistic optimization.

Furthermore, while \citet{ribeiro2023regularization} focused on constraining the sample-wise norm $\stdnorm{\bs \Delta}_{\alpha,\beta}$, \citet{xu2010robust} considered the feature-wise norm $\stdnorm{\bs \Delta^{\top}}_{\beta,\alpha}$. According to Theorem~1 of \citet{xu2010robust}, minimizing the pessimistic loss function $\max_{\stdnorm{\bs \Delta^{\top}}_{2,\infty} \le \tau} \stdnorm{\bs y-\bft(\bs X)}_2^2$ is equivalent to the square-root Lasso~\citep{belloni2011square} that minimizes the penalized loss function $\{\stdnorm{\bs y-\bs X\bs \theta}_2+\eta\stdnorm{\bs \theta}_1\}^2$ for some constant $\eta=\eta(\tau)$. The square-root Lasso is known for its theoretically favorable properties outlined by \citet{belloni2011square}. Parameter regularization fundamentally aims to stabilize estimation by reducing variance, thus enhancing the reliability of predictions. Therefore, increasing the pessimism (i.e., increasing $\tau$ in the pessimistic loss functions) is expected to diminish the variance of the estimator.

\subsubsection*{Outcome-pessimistic regression}

While covariate-pessimistic regression has a similar effect to regularization of the regression model, to the best of the author's knowledge, there is no positive evidence supporting the superiority of outcome-pessimistic regression. Furthermore, the defective nature of outcome-pessimism can be explained simply as follows. Using the rationale applied to covariate-pessimistic regression, we derive the following approximation for outcome-pessimistic regression (with the assumption $\alpha=2$ for simplicity):
\[
    \outop^{\pes}_{\alpha,\gamma,\tau}(\bs y,\bft(\bs X))
    \, \approx \,
    \left\{\stdnorm{\bs y-\bft(\bs X)}_2+\tau^{1/\gamma}\right\}^2-\tau^{2/\gamma}.
\]
This approximation implies that minimizing the outcome-pessimistic loss function closely resembles minimizing the straightforward OLS loss function $\stdnorm{\bs y-\bft(\bs X)}_2^2$. However, introducing outcome pessimism complicates the optimization process by potentially adding unnecessary variance (and in practice, bias) to the model. This complexity undermines the practical utility of outcome-pessimistic regression. Therefore, in contrast to the covariate-pessimistic regression which corresponds to introducing penalization for variations in the regression model $f_{\bs \theta}$, there is no evidence to support the superiority of outcome-pessimistic regression over OLS.

\section{Integration of Optimistic Regression and Pessimistic Regression}
\label{sec:experiments}

This section explores the bias-variance trade-off that appears in the optimism-pessimism continuum. Specifically, we consider outcome-optimism and covariate-pessimism, which are often employed practically and typically viewed as outlier-robust estimation and robust optimization, respectively. We do not consider covariate-optimism and outcome-pessimism, as the former is usually not considered a robust estimation method in the field of statistics, and there is no positive evidence to support the superiority of the latter.

To facilitate the exploration of the bias-variance trade-off, we define a loss function parameterized by a real-valued hyperparameter $\lambda \in \mathbb{R}$, which continuously bridges the outcome-optimistic and covariate-pessimistic loss functions:
\[
    \bridge_{\lambda}(\bs y,\bft(\bs X))
    =
    \begin{cases} 
    \outop^{\opt}_{1,1,25|\lambda|}(\bs y,\bft(\bs X)) & (\lambda > 0) \\
    \stdnorm{\bs y-\bft(\bs X)}_2^2 & (\lambda=0) \\
    \covop^{\pes}_{\infty,1,1,|\lambda|}(\bs y,\bft(\bs X)) & (\lambda < 0)
    \end{cases}. 
\]
Using these loss functions, we define the integrated estimator: 
\begin{align}
    \wht_{\lambda}
    :=
    \argmin_{\bs \theta \in \mathbb{R}^d} \bridge_{\lambda}(\bs y,\bft(\bs X)).
\label{eq:bridge_estimator}
\end{align}
Obviously, the estimator $\wht_{\lambda}$ is continuous with respect to $\lambda \in \mathbb{R}$, and it smoothly bridges the optimistic estimator ($\lambda > 0$) and pessimistic estimator ($\lambda < 0$) through the OLS estimator ($\lambda=0$). Specifically, increasing $\lambda$ enhances the optimism while decreasing $\lambda<0$ enhances the pessimism. We empirically demonstrate the bias-variance trade-off between the optimistic regression (i.e., outlier-robust estimation) and pessimistic regression (i.e., robust optimization) through the estimator \eqref{eq:bridge_estimator}.

\subsection{Bias and variance terms in generalization error}

This section describes the experimental settings and evaluation metrics. Source codes to reproduce the experimental results herein are available at \url{https://github.com/oknakfm/RBVT}. 

Consider a pair of random variables $(\bs x,y) \in \mathbb{R}^d \times \mathbb{R}$, where $\bs x \in \mathbb{R}^d$ is a $d$-dimensional covariate and $y \in \mathbb{R}$ is a scalar outcome. For simplicity, the conditional distribution of $y$ given $\bs x$ is modeled by a normal distribution $N(f_*(\bs x),\sigma_Y^2)$ with an underlying regression model $f_*$ and a standard deviation $\sigma_Y>0$. In most experiments, $f_*$ is specified as a linear model with the true coefficient $\bs \theta_*=(0,3/2,2/3)$, but later it will be specified as non-linear models in the discussion in Section~\ref{subsec:discussion}. 

To investigate the properties of the linear regression estimators with the setup described above, we utilize $n$ independent and identically distributed copies of $(\bs x,y)$, denoted as $\{(\bs x_i^{(t)},y_i^{(t)})\}_{i=1}^{n}$, to form a training dataset for each of $T$ trials. Here, $t=1,2,\ldots,T$ indicates the trial number. For an estimator $\wht^{(t)}=\wht(\bs X^{(t)},\bs y^{(t)})$ obtained using the training dataset at trial $t$, we evaluate the generalization error $T^{-1}\sum_{t=1}^{T} \mathbb{E}_{\bs x,y}\left[ \left\{ y - \mu^{(t)}(\bs x) \right\}^2 \right]$ over $T$ trials. Here, $\mu^{(t)}(\bs x)=\langle \wht^{(t)},\bs x\rangle$ denotes the estimated linear model at the trial $t$, and $\mu_T(\bs x):=T^{-1}\sum_{t=1}^{T} \mu^{(t)}(\bs x)$ denotes its average. Then, the generalization error can be decomposed into the sum of three terms $\sigma_Y^2,B_T^2,V_T$, where 
\begin{align*}
B_T := \left\{ \mathbbm{E}_{\bs x}\left[
    \left\{ f_{\bs \theta_*}(\bs x) - \mu_T(\bs x)\right\}^2
\right]\right\}^{1/2} \, \text{ and } \,
V_T := \frac{1}{T} \sum_{t=1}^{T} \mathbbm{E}_{\bs x}\left[
    \left\{\mu_T(\bs x) - \mu^{(t)}(\bs x) \right\}^2
\right]
\end{align*}
represent the bias and variance of the estimated regression function. Notably, $B_T$ and $V_T$ reduce to simpler forms $B_T=\sigma_X \cdot \norm{\bs \theta_*-\wbt_T}_2$ and $V_T=\sigma_X^2 \cdot T^{-1}\sum_{t=1}^{T}\norm{\wbt_T-\wht^{(t)}}_2^2$ if a linear function $f_*(\bs x)=\langle \bs \theta_*,\bs x \rangle$ with a true parameter $\bs \theta_*$ is employed and $\mathbb{E}_{\bs x}[\bs x\bs x^{\top}]=\sigma_X^2 \bs I_d$ is satisfied. 

In the next section, we will compute $B_T$ and $V_T$ for both optimistic and pessimistic regression methods, highlighting the bias-variance trade-offs between them.

\subsection{Bias-variance trade-off between optimistic and pessimistic linear regression}
\label{subsec:trade-off}

In this section, we consider the setting $d=3,\sigma_Y=1,\bs \theta_*=(0,3/2,2/3),T=100$. The covariate $\bs x$ is generated from a uniform distribution over $[-1,1]^d$. Once the dataset $\{(\bs x_i^{(t)},y_i^{(t)})\}_{i=1}^{n}$ for trial $t$ is generated, we incorporate artificial impediments: (a) covariate noise, where normal random numbers with mean $0$ and variance $0.5^2$ are added to all entries in the observed covariates $\bs x_i^{(t)}$; (b) outcome outliers, where the value $10$ is added to 5\% of randomly selected observed outcomes; (c) none, where the dataset is used as is for parameter estimation. 

\begin{figure}[!ht]
\centering
\begin{minipage}{0.32\textwidth}
\centering
\includegraphics[width=\textwidth]{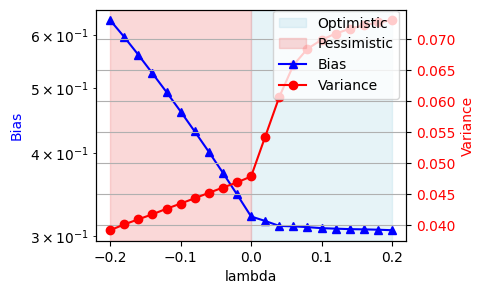}
\subcaption{Covariate noise}
\label{subfig:default_(a)}
\end{minipage}
\begin{minipage}{0.32\textwidth}
\centering
\includegraphics[width=\textwidth]{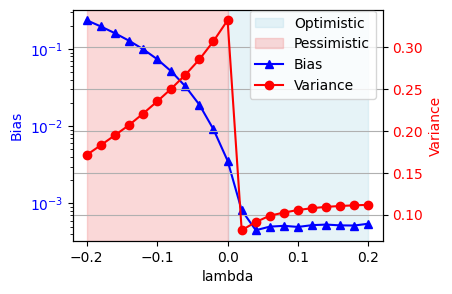}
\subcaption{Outcome outliers}
\label{subfig:default_(b)}
\end{minipage}
\begin{minipage}{0.32\textwidth}
\centering
\includegraphics[width=\textwidth]{figures/n50_d3_alpha1_sigma1_none.png}
\subcaption{None}
\label{subfig:default_(c)}
\end{minipage}
\caption{Default ($n=50, \sigma=1$)}
\label{fig:exp1_default}
\end{figure}

Using the datasets of three types (a)--(c), we compute the estimator \eqref{eq:bridge_estimator} by leveraging the \verb|cvxpy| package\footnote{\url{https://www.cvxpy.org/}} in \verb|Python| programming language\footnote{\url{https://www.python.org/}}, which effectively solves convex optimization problems. Subsequently, the bias $B_T$ and variance $V_T$ for the obtained estimators (for each $\lambda \in \mathbb{R}$) are computed. See Figure~\ref{fig:exp1_default} for the default setting ($n=50,\sigma=1$). 

\begin{figure}[!ht]
\centering
\begin{minipage}{0.32\textwidth}
\centering
\includegraphics[width=\textwidth]{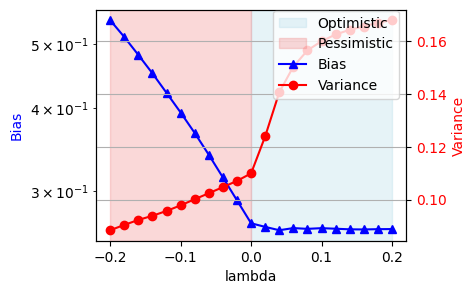}
\subcaption{Covariate noise}
\end{minipage}
\begin{minipage}{0.32\textwidth}
\centering
\includegraphics[width=\textwidth]{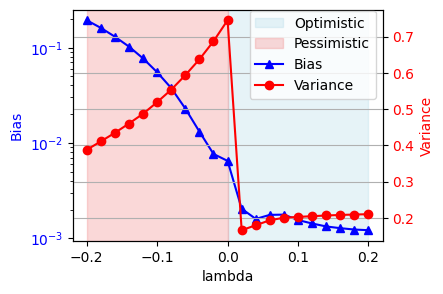}
\subcaption{Outcome outliers}
\end{minipage}
\begin{minipage}{0.32\textwidth}
\centering
\includegraphics[width=\textwidth]{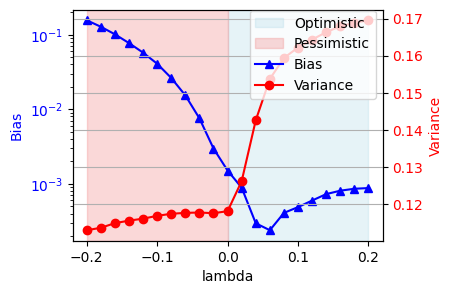}
\subcaption{None}
\label{subfig:smaller_n_(c)}
\end{minipage}
\caption{Smaller $n$ ($n=25, \sigma=1$)}
\label{fig:smaller_n}
\vspace{2em}
\centering
\begin{minipage}{0.32\textwidth}
\centering
\includegraphics[width=\textwidth]{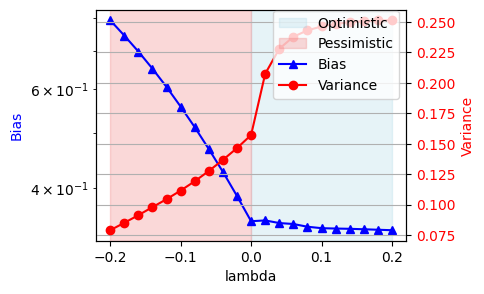}
\subcaption{Covariate noise}
\end{minipage}
\begin{minipage}{0.32\textwidth}
\centering
\includegraphics[width=\textwidth]{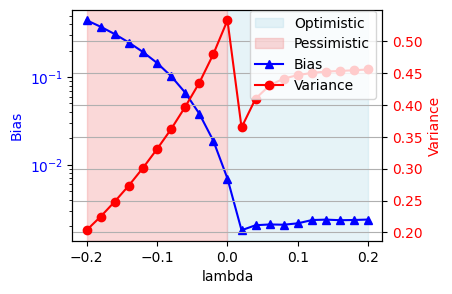}
\subcaption{Outcome outliers}
\end{minipage}
\begin{minipage}{0.32\textwidth}
\centering
\includegraphics[width=\textwidth]{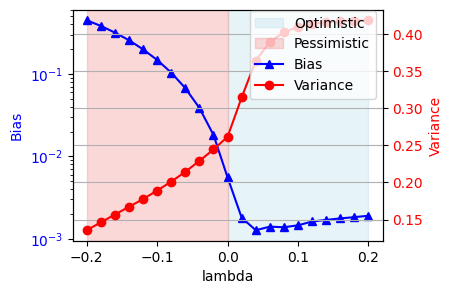}
\subcaption{None}
\label{subfig:larger_sigma_(c)}
\end{minipage}
\caption{Larger $\sigma$ ($n=50, \sigma=2$)}
\label{fig:larger_sigma}
\end{figure}

We can easily examine the bias-variance trade-off between outcome optimistic regression ($\eta > 0$) and covariate pessimistic regression ($\eta < 0$) in Figure~\ref{fig:exp1_default}(\subref{subfig:default_(a)}) and \ref{fig:exp1_default}(\subref{subfig:default_(c)}). Interestingly, for the outcome outliers as shown in Figure~\ref{fig:exp1_default}(\subref{subfig:default_(b)}), optimistic regression ($\eta > 0$) drastically reduces the variance if $\lambda$ is made slightly larger than $0$; it is because the adverse effect of outliers is removed by incorporating the outlier-robustness. This observation is consistent with Proposition~\ref{prop:gannaz}. For larger $\lambda$, variance increases (as well as other settings). The same tendency can be found for smaller $n=25$ shown in Figure~\ref{fig:smaller_n} and larger $\sigma=2$ shown in Figure~\ref{fig:larger_sigma}. 

Lastly, we note that the optimism with larger $\lambda>0$ can increase bias in some cases. For instance, see Figures~\ref{fig:smaller_n}(\subref{subfig:smaller_n_(c)}) and \ref{fig:larger_sigma}(\subref{subfig:larger_sigma_(c)}). This occurs because the OLS estimator is already good enough in this setup, and adding optimism introduces extra bias. 

Considering these observations, we provide further discussion in Section~\ref{subsec:discussion}. 

\subsection{Further discussion}
\label{subsec:discussion}

In this section, we discuss several important topics and share our insights into foreseen questions related to optimistic and pessimistic regression. 

\begin{enumerate}[{Q1.}]

\item \textbf{Is optimistic and pessimistic concept always effective? Does the solution space affect?} 
\label{enum:solution_space}

As discussed in \citet{hu2018does} (where \citet{hu2018does} considers DRO but not standard RO), sometimes optimization does not have a significant effect. For instance, similar to \citet{hu2018does}, we can easily prove that $\outop^{\opt}_{2,2,\tau}(\bs y,\bft(\bs X)),\outop^{\pes}_{2,2,\tau}(\bs y,\bft(\bs X)) \propto \stdnorm{\bs y-\bft(\bs X)}_2^2$ for any $\tau>0$. This indicates that optimistic and pessimistic regression using $2$-norm constraints lead to the same results as OLS. These facts suggest that the selection of the solution space is crucial for effectively utilizing the concepts of optimism and pessimism. However, to the best of the authors' knowledge, no existing study reveals the relationship between the estimator property and the type of constraints. Specifying the appropriate constraints remains a problem worth considering in future work.

\item \textbf{Which should be employed in practice: optimistic regression or pessimistic regression? Are there any recommendations?}
\label{enum:recommendation}

Although it depends on the specific situation, here we provide some insights. 

Firstly, when the regression model is correctly specified (where the true model and regression model are both linear, as in Section~\ref{sec:experiments}), there exists a simple bias-variance trade-off. Optimism and pessimism provide roughly converse strategies to reduce the generalization error, which is the sum of the squared bias and variance. As discussed, enhancing optimism reduces bias (but increases variance as compensation), whereas enhancing pessimism reduces variance (but increases bias as compensation).

Secondly, the strengths of optimism and pessimism differ. Optimism (i.e., outlier-robustness) can drastically reduce variance if the outcomes include several outliers. Therefore, even if there is still a bias-variance trade-off within optimistic regression (as $\lambda>0$ increases), the equilibrium point of bias and variance can shift towards lower values (only in the region $\lambda>0$), thereby reducing the total generalization error even compared to pessimistic regression. However, if the regression model is misspecified, optimistic regression can increase both bias and variance. In such cases, pessimistic regression can reduce the overall generalization error (depending on the settings), as discussed in the next \ref{enum:misspecification}.

\item \textbf{Are optimistic and pessimistic methods also robust against model misspecification?}
\label{enum:misspecification}

In general, no. Specifically, optimistic regression (i.e., outlier-robust regression) suffers significantly from model misspecification. Figure~\ref{fig:model_misspecificaiton} illustrates the bias and variance of the linear regression model under model misspecification. In these experiments, we employ non-linear functions (a) $f_*(\bs x)=\exp(\langle \bs x,\bs \theta_*\rangle)$, (b) $f_*(\bs x)=\cos(\langle \bs x,\bs \theta_*\rangle)$, (c) $f_*(\bs x)=\{\langle \bs x,\bs \theta_*\rangle\}^2-1$. 

\begin{figure}[!ht]
\centering
\begin{minipage}{0.32\textwidth}
\centering
\includegraphics[width=\textwidth]{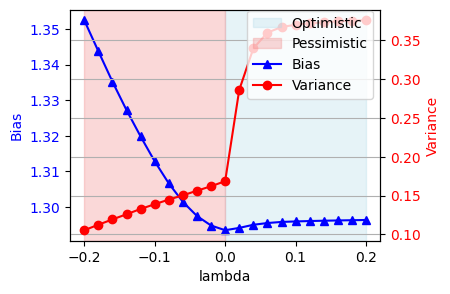}
\subcaption{$f_*(\bs x)=\exp(\langle \bs x,\bs \theta_*\rangle)$}
\label{subfig:exp}
\end{minipage}
\begin{minipage}{0.32\textwidth}
\centering
\includegraphics[width=\textwidth]{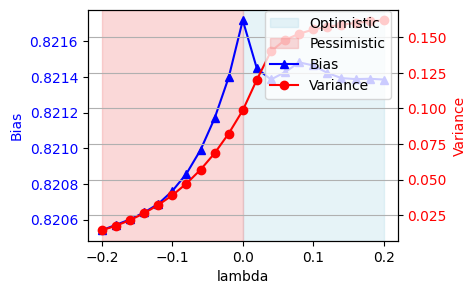}
\subcaption{$f_*(\bs x)=\cos(\langle \bs x,\bs \theta_*\rangle)$}
\label{subfig:quad}
\end{minipage}
\begin{minipage}{0.32\textwidth}
\centering
\includegraphics[width=\textwidth]{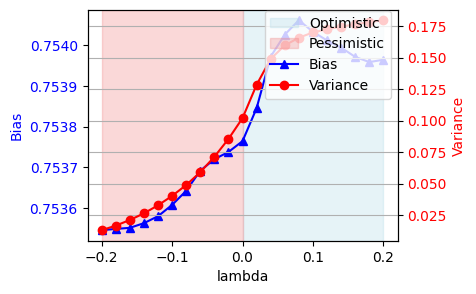}
\subcaption{$f_*(\bs x)=\{\langle \bs x,\bs \theta_*\rangle\}^2-1$}
\label{subfig:quad-1}
\end{minipage}
\caption{Model misspecification (neither covariate noise nor outcome outlier are included)}
\label{fig:model_misspecificaiton}
\end{figure}

Optimistic regression can increase both bias and variance under model misspecification, while it generally reduces bias when the model is correctly specified, as shown in Section~\ref{sec:experiments}. Therefore, under model misspecification, there is no strong evidence supporting the utility of optimistic methods.

On the other hand, with pessimistic regression (i.e., robust optimization), we observe that variance almost always decreases as pessimism is enhanced (at least in the above experiments). Whether bias decreases or increases as pessimism is enhanced depends on the problem setting, but pessimistic regression at least decreases variance. Therefore, regardless of whether the regression model is correctly specified, we may suggest introducing a small degree of pessimism. The introduction of a small degree of pessimism demonstrates good properties in asymptotic theory ($\lambda=\lambda_n \to 0$), as shown in \citet{imaizumi2023sup}.

\item \textbf{Does the pessimistic method always reduce the variance?} 
\label{enum:variance_in_pessimistic}

No. For instance, if OLS is already sufficient (e.g., when the regression model is correctly specified and the sample size $n$ is large enough), OLS can achieve the smallest bias and variance. See Figure~\ref{fig:OLS_is_the_best} for this situation. In such cases, additional operations to enhance optimism or pessimism can introduce extra bias and variance. However, in practice, such ideal situations rarely occur. Also, introducing small dataset perturbations (as \citet{imaizumi2023sup} did in their asymptotic theory) does not conflict with the effectiveness of OLS. Therefore, we conclude that a pessimistic approach is effective for general problems, as long as the solution space is appropriately and severely constrained. This conclusion again highlights the importance of the influence function~\citep{hampel1986robust}, which measures the estimator's sensitivity to the impediments.

\begin{figure}[!ht]
\centering
\includegraphics[width=0.33\textwidth]{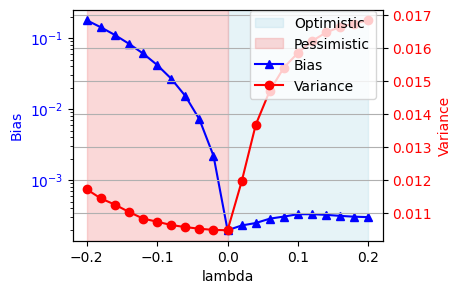}
\caption{Correctly specified with sufficient sample size $n=250$, where the dataset does not include either covariate noise nor outcome outlier (i.e., OLS is already good enough).}
\label{fig:OLS_is_the_best}
\end{figure}

\item \textbf{Is there a similar attempt that bridges optimistic and pessimistic optimization?} 
\label{enum:bridge}

Although it is outside the normal use and not exactly the same, we think that density power divergence~\citep[DPD;][]{basu1998robust,basu2011statistical}, which can robustly measure the discrepancy between two distributions against outliers, has a similar property. This property is achieved by placing the maximum likelihood estimator between the positive exponent $\beta>0$ (corresponding to optimism) and the negative exponent $\beta<0$ (corresponding to pessimism). To explain this similarity, we consider an estimating equation for DPD applied to a regression problem:
\begin{align*}
\bs 0 
\, = \, 
\frac{\partial}{\partial \bs \theta}&
\frac{1}{n}\sum_{i=1}^{n} D_{\beta}(\widehat{Q}_i(\cdot \mid \bs x_i),P_{\theta}(\cdot \mid \bs x_i)) \\
&= \, 
-
\frac{1}{n}\sum_{i=1}^{n}
\int
p_{\bs \theta}(y \mid \bs x_i)^{\beta}
\{\hat{q}_i(y \mid \bs x_i) - p_{\bs \theta}(y \mid \bs x_i)\}
\frac{\partial \log p_{\bs \theta}(y \mid \bs x_i)}{\partial \bs \theta}
\diff y,
\end{align*}
where $\hat{q}_i$ denotes the empirical density, i.e., the density function of the empirical distribution $\hat{Q}_i(y \mid \bs x_i)=\mathbbm{1}(y \le y_i)$. $\beta=0$ corresponds to the simple estimating equation for the likelihood-based regression, and $\beta=-1$ is especially called a Itakura-Saito divergence~\citep{itakura1968analysis}.

As we can easily prove, if the exponent $\beta$ in DPD is positive, each term in the estimating equation is lightly weighted if $p_{\bs \theta}(y \mid \bs x_i)$ is small, i.e., $y$ is an outlier. Therefore, DPD with positive $\beta>0$ is known to yield outlier-resistant property (as outliers are lightly weighted), corresponding to optimism. On the other hand, if the exponent $\beta$ is negative, the term is highly weighted if $y$ is an outlier. 

\begin{figure}[!ht]
\centering
\begin{minipage}{0.32\textwidth}
\centering
\includegraphics[width=\textwidth]{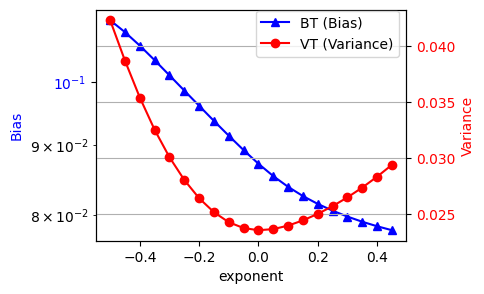}
\subcaption{Covariate noise}
\end{minipage}
\begin{minipage}{0.32\textwidth}
\centering
\includegraphics[width=\textwidth]{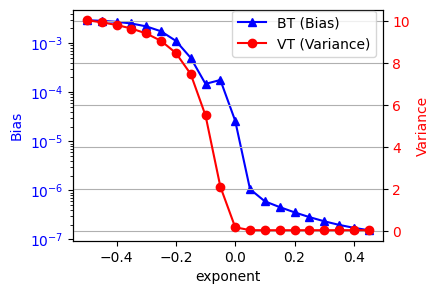}
\subcaption{Outcome outliers}
\end{minipage}
\begin{minipage}{0.32\textwidth}
\centering
\includegraphics[width=\textwidth]{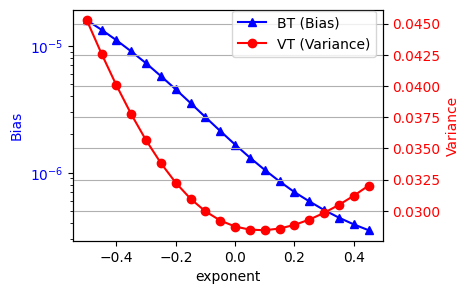}
\subcaption{None}
\end{minipage}
\caption{DPD regression}
\label{fig:DPD_regression}
\end{figure}

However, note that DPD with a negative exponent $\beta<0$ does not exactly correspond to pessimism. See Figure~\ref{fig:DPD_regression}, which shows the bias $B_T$ and variance $V_T$ for the estimator obtained by minimizing the density power divergence with a given standard deviation $\sigma_Y$. While the overall tendency for optimism ($\beta>0$ for DPD and $\lambda>$ for the estimator \eqref{eq:bridge_estimator}) is the same as in previous experiments, DPD regression differs from pessimistic regression. 

\item \textbf{Is there any attempt that mixes optimistic and pessimistic method to reduce both bias and variance?}

\citet{hashimoto2018fairness} pointed out that distributionally robust optimization can be sensitive to outliers, raising the open problem of whether outlier-robustness can be appropriately incorporated into distributional robustness. Some subsequent studies have attempted to address this question: \citet{zhang2019theoretically} considers a robust regularization term expected to reduce the sensitivity of the estimated functions, while \citet{zhai2021DORO} defines a specific dataset distribution perturbation representing the contaminated outlier model~\citep{huber2011robust} and conducts distributionally robust optimization using the outlier-based solution space. 
 
\end{enumerate}

\section{Conclusion}
\label{sec:conclusion}

In this paper, we considered an integrated perspective on robustness in regression. More specifically, we described the relationship between outlier-resistant robust estimation (optimism) and robust optimization (pessimism). Through numerical experiments, we demonstrated the bias-variance trade-off between optimism and pessimism and provided several discussions. Source codes to reproduce the experimental results are available at \url{https://github.com/oknakfm/RBVT}. 

Regarding bias and variance, our conclusions are summarized as follows:

\begin{itemize}
\item Variance in the optimistic method generally increases. However, if the model is correctly specified and outliers are included in the outcomes, the variance drastically decreases compared to OLS by incorporating a small amount of optimism, as shown in Section~\ref{subsec:trade-off}. We note that even in this setting, the variance increases as the level of optimism is enhanced, indicating that the bias-variance trade-off still exists in the region $\lambda>0$.

\item Bias in the optimistic method generally decreases, as long as the regression model is correctly specified (see Section~\ref{subsec:trade-off}). However, if the model is misspecified, the bias can increase, as discussed in \ref{enum:misspecification} in Section~\ref{subsec:discussion}.

\item Variance in the pessimistic method generally decreases. However, in ideal situations (e.g., the regression model is correctly specified and the sample size $n$ is large enough), OLS already achieves a good estimator. In such cases, introducing pessimism may cause extra variance, as discussed in \ref{enum:misspecification} in Section~\ref{subsec:discussion}.

\item Bias in pessimistic method is increased in general as long as the regression model is correctly specified (see Section~\ref{subsec:trade-off}). However, in the case that the regression model is misspecified, the bias reduces depending on the true regression function $f_*(\bs x)$ as discussed in \ref{enum:misspecification} in Section~\ref{subsec:discussion}. Practically speaking, it is hard to preliminarily distinguish whether the pessimistic method applied to a specific dataset may increase the bias or not. 
\end{itemize}

\section*{Acknowledgement}
We are grateful to Han Bao, Takayuki Kawashima, Shotaro Yagishita, and Hironori Fujisawa for their insightful discussions. A. Okuno was supported by JSPS KAKENHI Grants (21K17718, 22H05106).

\bibliographystyle{apalike}
\bibliography{rbvt}

\end{document}